\documentclass[manuscript,screen]{acmart}
\AtBeginDocument{%
  \providecommand\BibTeX{{%
    \normalfont B\kern-0.5em{\scshape i\kern-0.25em b}\kern-0.8em\TeX}}}

\copyrightyear{2023}
\acmYear{2023}
\setcopyright{acmlicensed}\acmConference[CNIOT '23]{2023 4th International
Conference on Computing, Networks and Internet of Things}{May 26--28,
2023}{Xiamen, China}
\acmBooktitle{2023 4th International Conference on Computing, Networks and
Internet of Things (CNIOT '23), May 26--28, 2023, Xiamen, China}
\acmPrice{15.00}
\acmDOI{10.1145/3603781.3603869}
\acmISBN{979-8-4007-0070-5/23/05}

\begin{document}

\title{A Novel Scholar Embedding Model for Interdisciplinary Collaboration}

\author{Yitong Hu}\authornote{All authors contributed equally to this research.}
\email{huyt@bupt.edu.cn}
\orcid{0009-0008-1697-1465}

\author{Zixuan Zhu}\authornotemark[1]
\email{ternura@bupt.edu.cn}

\author{Yizhe Wang}\authornotemark[1]
\email{anonymity@bupt.edu.cn}

\author{Junxiang Wang}
\email{1034696975@bupt.edu.cn}

\author{Zehao Xing}
\email{sonaldovski@bupt.edu.cn}

\affiliation{%
  \institution{Beijing University of Posts and Telecommunications}
  \streetaddress{No. 10, Xitucheng Road}
  \city{Beijing}
  \country{China}
  \postcode{100876}
}

\renewcommand{\shortauthors}{Hu, Zhu and Wang, et al.}

\begin{abstract}
Interdisciplinary collaboration has become a driving force for scientific breakthroughs, and evaluating scholars' performance in interdisciplinary researches is essential for promoting such collaborations. However, traditional scholar evaluation methods based solely on individual achievements do not consider interdisciplinary cooperation, creating a challenge for interdisciplinary scholar evaluation and recommendation. To address this issue, we propose a scholar embedding model that quantifies and represents scholars based on global semantic information and social influence, enabling real-time tracking of scholars' research trends. Our model incorporates semantic information and social influence for interdisciplinary scholar evaluation, laying the foundation for future interdisciplinary collaboration discovery and recommendation projects. We demonstrate the effectiveness of our model on a sample of scholars from the Beijing University of Posts and Telecommunications.
\end{abstract}





\maketitle

\hypertarget{introduction}{%
\section{Introduction}\label{introduction}}

The intersection of disciplines is often the new growth point of science
and the new scientific frontier, where major scientific breakthroughs
are most likely to occur and revolutionary changes in science. We need
more scholars to collaborate across disciplines to promote the development
of science. Therefore, we need a systematic discovery and recommendation
solution of interdisciplinary cooperation network, that is, to discover
and evaluate the interdisciplinary cooperation among scholars, and
recommend suitable cooperation objects according to the needs of the scholar,
so as to promote interdisciplinary cooperation among scholars. 
The premise of studying the above problems is to complete the scholar
evaluation under the interdisciplinary background.

The traditional scholar evaluation refers to the quantitative evaluation
of the academic level and influence of the scholar. Common indicators include
the number of papers published, the citation frequency of papers, the
citation frequency of papers per citation, the number of highly cited
papers and H index, etc. However, the above methods are all based on the
evaluation of individual achievements of the scholar and do not take into
account the cooperation of the interdisciplinary scholar, which brings
challenges to the scholar evaluation and cooperative recommendation in
the era of interdisciplinary research.

It is worth noting that different subjects and fields may have
various scholar evaluation indicators and methods. Although the
traditional method based on the characteristics of the subject is
accurate, it requires a huge amount of manpower and material resources,
and it has a considerable lag. With the rapid development of
interdisciplinary research, the traditional scholar assessment based on the
characteristics of the subject can not adapt to the features of 
interdisciplinary research, including numerous, more subdivisions, 
complex structure, and rapid development. Specifically, almost every year,
multiple interdisciplinary studies are born, which combine
characteristics from different subjects and may undergo structural
changes due to breakthrough results in one subject. Because of their
own limitations, scholar assessment designers are difficult to develop
detailed assessment methods for the lack of history, wide span and rapid
development of interdisciplinary. These facts pose a great challenge to
the interdisciplinary scholar evaluation: to improve the real-time,
universal (interdisciplinary) and self-adaptability of the evaluation methods.

\begin{figure}[htbp]
  \centering
  \includegraphics[width=\linewidth]{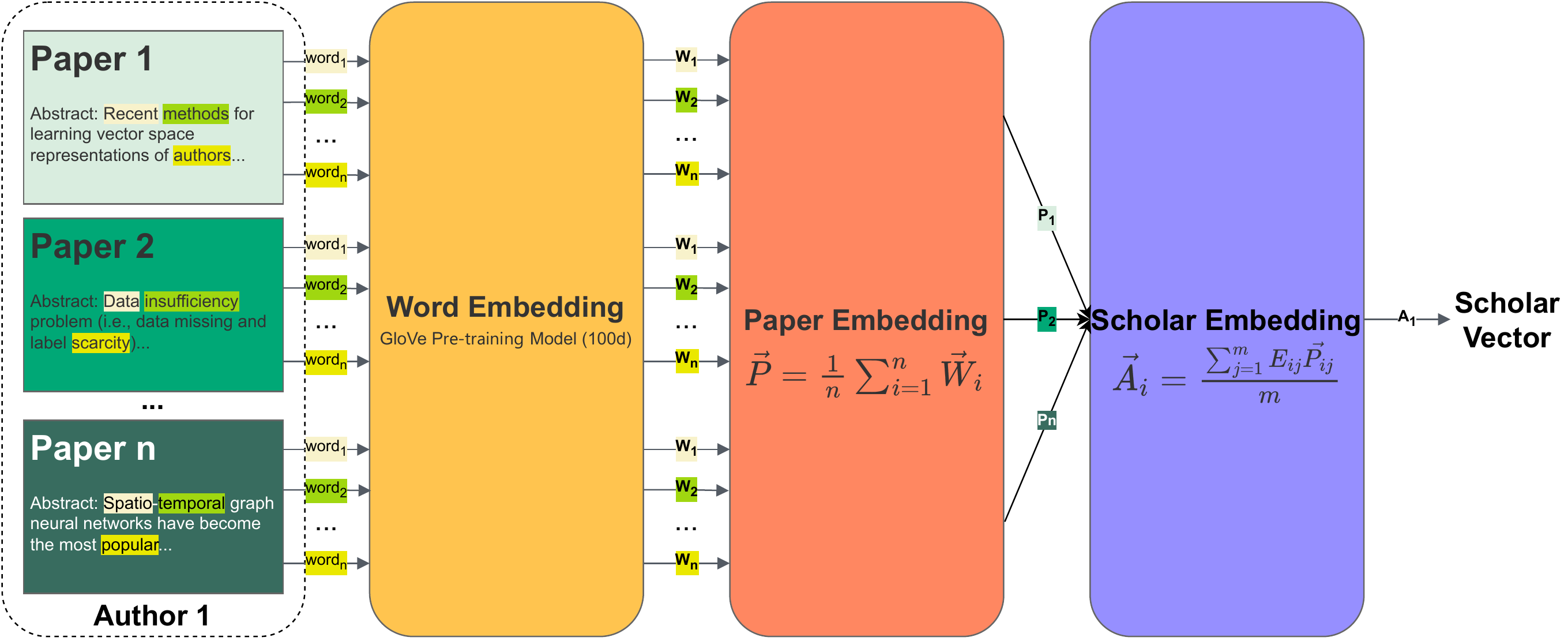}
  \caption{Introduction of Scholar Embedding Model.}
  \Description{Dimensionality Reduction Transformation.}
  \label{fig:intro}
\end{figure}

To tackle the challenges, we propose an evaluation paradigm for
the interdisciplinary scholar. As we know, the necessary work before evaluation is
to choose appropriate methods to quantify the academic achievements of
the scholar in various fields. Therefore, this paper focuses on the
quantification and representation of the scholar, that is, embedding the scholar
into a lower-dimensional continuous vector space. Specifically, we
propose an innovative global semantic information and social influence based
scholar embedding model for scholar representation in the form of vector, as shown in Figure \ref{fig:intro}. The
input of our model consists of global semantic information and
social influence. Among them, semantic input refers to
the abstract of all the papers of the scholar, which represents all the
academic achievements of the scholar in various subjects. 
The introduction of semantic input allows us to break through barriers between different subjects,
and assess the interdisciplinary scholar in a more fine-grained semantic dimension with universal applicability.
Social influence input is a comprehensive measure of the scholar's influence and
contribution in various disciplines. It reflects the scholar's expertise and
cooperation in interdisciplinary studies. At the same time, the two inputs above
are calculated in real-time according to the data of paper abstract,
paper publication volume, paper citation frequency, etc., which enables
us to track the research trends of the scholar in real-time.

\newpage
The contributions of this paper are summarized as follows:
\begin{itemize}
\item
  To the best of our knowledge, our scholar embedding model is the first
  quantification and presentation model for cross-disciplinary scholar.
  It takes into account the paper of scholar as well as their global
  semantic information and social influence, and is a real-time
  evaluation for scholar of cross-disciplinary. Solutions to universal
  and adaptive challenges.
\item
  Discovery and recommendation of interdisciplinary cooperation network
  is a complex system project, and our scholar embedding model has solved
  the fundamental work of scholar quantification and presentation, laying
  a foundation for future work, including cross scholar assessment,
  discovery, evaluation and recommendation of interdisciplinary
  cooperation network, and more innovative applications.
\item
 We take 126 scholars from Beijing University of Posts and Telecommunications (BUPT) as a showcase
  and empirical demonstration the sentiment of our
  proposed scholar embedding model on scholar similarity task. Moreover,
  our work provides data-supported insights into the interdisciplinary
  construction of BUPT.
\end{itemize}

\hypertarget{related work}{%
\section{Related Work}\label{related work}}

\hypertarget{word2vec}{%
\subsection{Word2vec}\label{word2vec}}

In order to carry out quantitative analysis of the paper abstract of
various author, words need to be mapped to the vector with word
characteristics, namely word embedding technology. Word embedding is the
vector representation of a word, which can capture its semantic and
syntactic meaning. While word2vec \cite{7389336} maps each word to a
fixed-length vector that better expresses the similarity \cite{https://doi.org/10.48550/arxiv.2204.04833} and analogy
relationship between different words. Their training relies on their own
conditional probabilities. Under the assumption of the bag-of-words
model in word2vec, the order of words is not important. The calculation
process of word2vec is to iterate over all the training data, and
finally we can get that words with similar contexts have similar
semantics, and the cosine similarity of the word vectors corresponding
to these words will also get high values.

The word2vec tool consists of two main models \cite{7259377}: skip-grams and
continuous bag of words (CBOW), which is closely related to
GloVe. The hop metamodel assumes that a word can be used to generate its
surrounding words in a text sequence. In the hop metamodel, the
parameters are the vector of the head word and the vector of the context
word for each word in the vocabulary, each of which is represented by
two \(d\)dimensional vectors for computing the conditional probability.
For the context window \(m\), the likelihood function of the jump
meta-model is the probability of generating all the context words given
any head word. This is not the end of the story, because if you just add
a Softmax \cite{8852459} activation function, the calculation is still large, there
will be as many dimensions as there are words. This is why we propose
Hierarchical Softmax, which uses Huffman Tree to encode the lexicon of
the output layer, instantly reducing the dimensionality to the depth of
the tree. However, the problems existing in word2vec are that each local
context window is trained separately, the statistical information
contained in the global co-occurrence matrix is not utilized, and the
polysemous words cannot be well represented and processed because the
unique word vector is used.

\hypertarget{glove}{%
\subsection{GloVe}\label{glove}}

GloVe \cite{GloVe} is an extension of the word2vec method for
generating word embeddings, which attempts to improve some of the
limitations of the original word2vec method. Although both word2vec and
GloVe generate word embeddings \cite{https://doi.org/10.48550/arxiv.2301.10656} by training on a large amount of text
data, GloVe takes a different approach to model the relationships
between words. Specifically, GloVe attempts to capture co-occurrence
statistics between words in the corpus, rather than capturing only the
local context around each word as word2vec does. By doing so, GloVe is
able to capture both global and local relationships between words,
resulting in embeddings that are more suitable for capturing semantic
relationships between words \cite{7478417}.

However, in GloVe, the objects being embedded are words, not papers
or authors, which is different from the problem we are studying.
Therefore we need to use GloVe for some sequence-level applications to
propose new embedding methods that can be applied to our problem.

\hypertarget{preliminaries}{%
\section{Preliminaries}\label{preliminaries}}

Word embeddings, also known as distributed representations, are widely
used techniques in natural language processing (NLP) that enable
machines to capture the meaning and context of words. Global Vectors for
Word Representation (GloVe) is one such approach, which has gained
popularity for its ability to produce high-quality word embeddings \cite{8861084}.

GloVe is introduced in a paper by Pennington and others, which proposes a new method for training word
embeddings based on the global co-occurrence statistics of words in a
corpus. The method involves decomposing the word co-occurrence count
matrix using singular value decomposition (SVD) and learning word
vectors that capture the statistical relationships between words.

The key formulation used in GloVe is the objective function:
\begin{equation}\label{eq:J}
    J = \sum_{i,j=1}^{V}f(X_{ij})(w_i^T \tilde{w_j} + b_i + \tilde{b_j} - log(X_{ij}))^2
\end{equation}
where \(V\) is the vocabulary of words, \(X_{ij}\) is the number of
co-occurrences of words \(i\) and \(j\), \(w_i\) and \(w_j\) are word
vectors, \(b_i\) and \(b_j\) are biased words, and \(f(X_{ij})\) is a
weighting function that assigns smaller weights to rare word pairs. The
goal is to minimize \(J\) by adjusting the word vectors and biases.

Another important formula in word embedding is cosine similarity:
\begin{equation}\label{eq:similarity}
    similarity(w_i, w_j) = cos(\theta) = \frac{w_i^T * w_j}{||w_i|| * ||w_j||}
\end{equation}

It measures the similarity between two word vectors \(w_i\) and \(w_j\)
by calculating the cosine of the angle between them. A high cosine
similarity value indicates that the words are semantically similar.
Overall, word embedding represented by GloVe has become an integral part
of NLP and has enabled many applications such as machine translation,
sentiment analysis, and text classification.

\hypertarget{scholar embedding model}{%
\section{Scholar Embedding Model}\label{scholar embedding model}}

When applying GloVe to the problem of this article, i.e., constructing a
semantic and combined influence-based cooperative network, the
corresponding principle is to transform the paper summaries of
scholars into vectors using GloVe's pre-training model, a process called
paper embedding. Then, based on the combined influence of each
paper, we perform a weighted average operation on these paper
vectors to obtain the scholar vectors, a process called scholar
embedding. Finally, we construct an interdisciplinary academic
collaboration network based on semantics and combined influence based on
the scholar vector. One advantage of using GloVe to process this work is
that the embedding captures the semantic and syntactic relationships
between words, which may be important for identifying meaningful
research connections between scholars across disciplines. In addition,
GloVe embeddings have been shown to perform well in a variety of natural
language processing tasks, suggesting that they may be a powerful tool
for identifying interdisciplinary collaborations.

\hypertarget{paper embedding}{%
\subsection{Paper Embedding}\label{paper embedding}}

We define the vector \(\vec P\) of the paper:
\begin{equation}\label{eq:P}
    \vec P = \frac{1}{n}\sum_{i=1}^{n} \vec W_i
\end{equation}
where \(\vec W_i\) is the word vector of the \(i\)th word in the
paper abstract and \(n\) is the total number of words in the
paper abstract. Here we use the mean value to represent the
paper vector, because we believe that the topic of the paper
should be determined by all words in the abstract together.

For the word vector \(\vec W\), we take the paper abstracts in the
database (data type: String), slice them into words, and then map each
word (token), into GloVe's pre-trained model to obtain a
100-dimensional word vector. In this paper, we use GloVe's
pre-trained model (6B tokens, 400K vocab, uncased, 100d vectors), which
is trained on the Wikipedia corpus.

\hypertarget{scholar embedding}{%
\subsection{Scholar Embedding}\label{scholar embedding}}

\begin{figure}[htbp]
  \centering
  \includegraphics[width=\linewidth]{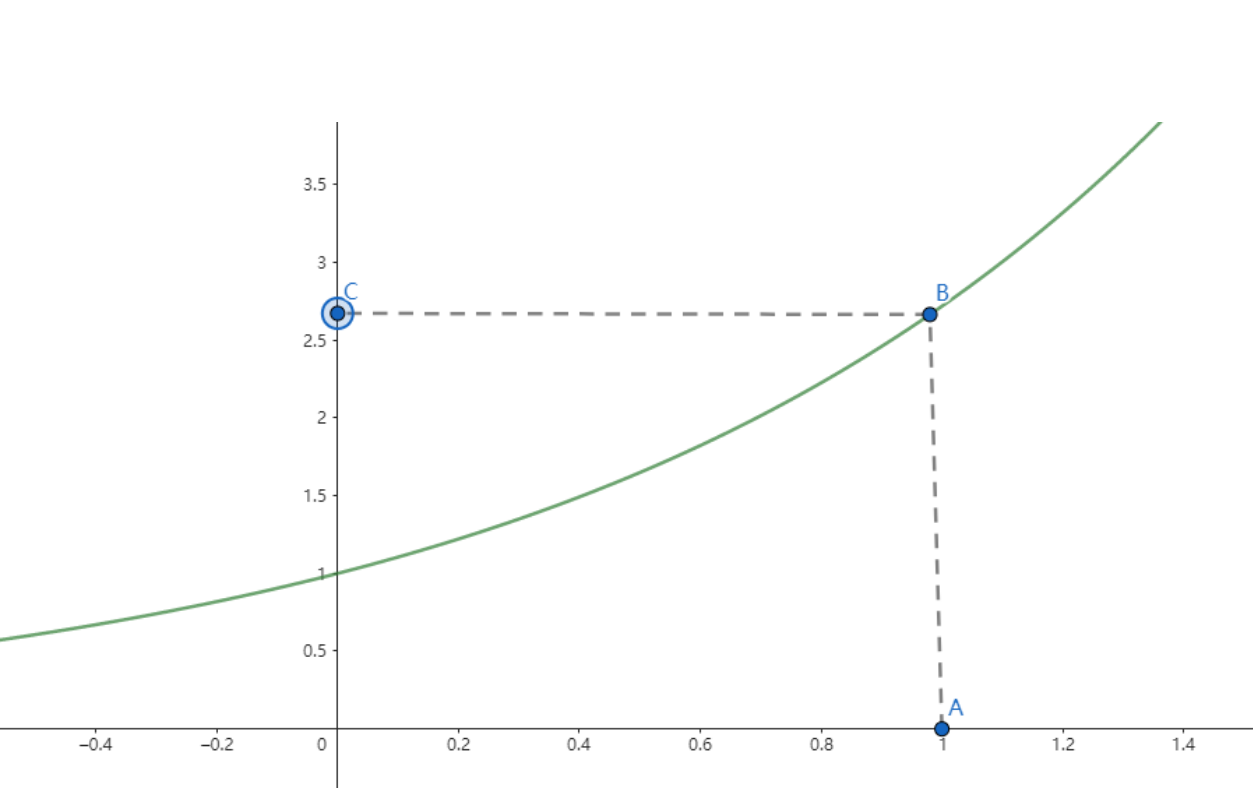}
  \caption{The Exponential Function of the Interval [0, 1].}
  \Description{The Exponential Function of the Interval [0, 1].}
  \label{fig:exponential}
\end{figure}

First, we define some notation. \(x_{ij}\) represents the total number of
co-authors in the paper \(j\) of scholar \(i\). In general, for a
paper, the higher the scholar's ranking, the greater his
contribution to that paper. Therefore, we can assume that the scholar's contribution is non-uniformly proportional to his ranking in that paper. An exponential function shown in Figure \ref{fig:exponential} is used to represent the scholar's contribution.

Based on the above assumptions, the contribution of scholar \(i\) to the
paper \(j\), \(E_{Rij}\), is expressed as follows.
\begin{equation}\label{eq:Er}
    E_{Rij}= \begin{cases} e,& \text{k = 1} \\ e^{\frac{x_{ij}-k}{x_{ij}}}, & \text{others} \end{cases}
\end{equation}
where \(k\) denotes the rank of the scholar among all co-authors in the
paper.

In addition, the number of citations of that paper reflects the
influence of that paper, and usually, the more citations, the
higher the influence of that paper. Therefore, we define the impact
factor \(C_{ij}\) for the paper \(j\) of scholar \(i\) as follows.
\begin{equation}\label{eq:C}
    C_{ij} = \begin{cases} 0,& {{n_{ij}} = 0} \\ 1, & {{n_{ij}} > 0} \end{cases}
\end{equation}
where \(n_{ij}\) is the number of times the paper is cited. And
combined with the impact factor \(C_{ij}\) of the paper, we can
obtain the impact \(E_{Cij}\) of the paper \(j\), which is defined
as follows.
\begin{equation}\label{eq:Ec}
    E_{Cij} = C_{ij} e^{\frac{n_{ij}}{n_{i_{max}}}}
\end{equation}
where \(n_{ij}\) is the number of citations of paper
\(j\), \(n_{i_{max}}\) is the highest single citation number of scholar \(i\).

The above two steps present the contribution of the scholar to the paper and the influence of the paper, respectively. Next, we improve the above equation based on the idea of PageRank \cite{ilprints422}, which considers the influence of other factors based on the influence of a single point of the network, and we adapt the model:
\begin{equation}\label{eq:E}
   E_{ij} = \lambda E_{Rij} + (1-\lambda)E_{Cij}
\end{equation}
\(E_{ij}\) represents the combined influence of paper \(j\), here
we consider two factors. \(E_{Rij}\), the contribution of scholar \(i\) to paper
\(j\). And \(E_{Cij}\), the influence of paper \(j\),
where the weight parameter \(\lambda \in [0, 1]\) determines the influence of both
\(E_{Rij}\) and \(E_{Cij}\) on \(E_{ij}\).

Through the above steps we obtain the influence \(E_{ij}\) of scholar
\(i\) in paper \(j\), and next multiply it with the vector
\(\vec{P_{ij}}\) of paper \(j\) of scholar \(i\) extracted by GloVe
and perform the weighted average operation to obtain the vector
\(\vec{A_i}\) of scholar \(i\), which is defined as follows.
\begin{equation}\label{eq:A}
    \vec{A_i} = \frac{\sum_{j=1}^{m} E_{ij} \vec{P_{ij}}}{m}
\end{equation}
where \(m\) denotes the total number of documents written by scholar
\(i\). By this method we can get a vector representation of each
scholar, and we call the above process Scholar Embedding.

\hypertarget{model complexity}{%
\subsection{Model Complexity}\label{model complexity}}

As can be seen from Equation \ref{eq:E}, the time complexity of calculating the influence of an author in an article depends on the time complexity of calculating 
\(E_{Rij}\) and \(E_{Cij}\). As can be seen from
Equation \ref{eq:Er}, \(E_{Rij}\) depends on the extent of the author's
contribution in this article (i.e., the authorship order of this
article). In this model \(E_{Rij}\) can be calculated by reading the
author order of the article in question and calling the formula
Equation \ref{eq:Er} to directly derive the contribution \(E_{Rij}\) of
this author under this article.

As can be seen from Equation \ref{eq:C}, \ref{eq:Ec},
\(E_{Cij}\) depends on the influence (number of citations) of the
article in the academic community. In this model \(E_{Cij}\) can be
calculated by reading the number of citations of the article in question
and calling the formula Equation \ref{eq:C} and the
Equation \ref{eq:C} to directly derive the impact \(E_{Cij}\) of the
article.

For the step of calculating the influence of an author on an article
using the author ranking and citation counts of the published paper
by authors after the GloVe calculation, the time complexity of this
model is: \(O(1)\) and is positively correlated with the number of
articles as follows.
\begin{equation}\label{eq:complexity}
    \mid n \mid \backsim O(1)
\end{equation}

Assuming that an author publishes a total of n articles, the overall
time complexity of this step is \(O(n)\).

In summary, under this model, the time complexity of converting the raw
data obtained from the ranking of authors and the number of citations,
etc. of all publications by scholars into the final total paper
impact of authors is (where \(O(|C|^{0.8})\) is the time complexity of
the GloVe algorithm and \(n\) is the total number of publications by
authors):
\begin{equation}\label{eq:complexity2}
    O(n \cdot |C|^{0.8})
\end{equation}

The model uses GloVe, which is currently the best performing word vector
processing tool in the word vector domain, and its time complexity of
\(O(|C|^{0.8})\) is better than the rest of the word vector processing
tools in terms of time complexity, which meets the time complexity
requirement of this project. For the step of converting the processed
word vectors into the final total scholarly paper impact, the model
reduces the time complexity of processing individual articles to
\(O(1)\) and the time complexity of processing single author total
paper impact to \(O(n)\), which performs well and the model meets
the requirements.

\hypertarget{experiments}{%
\section{Experiments}\label{experiments}}

\hypertarget{model testing methods}{%
\subsection{Model Testing Methods}\label{model testing methods}}

To evaluate the effect of scholar embedding, we conduct experiments on
the author similarity task. Pennington \cite{GloVe}
performed word similarity evaluation \cite{https://doi.org/10.48550/arxiv.2211.08203} on GloVe, and accordingly, we refer
to their work The scholar similarity task is designed because our
embedding is based on the GloVe pre-trained model. Specifically, we
invited experts from BUPT to evaluate the collaborative relationships of 126 scholars in our
database. The cooperative relationships are divided into 5 levels (1-5), corresponding to the values of scholar similarity of 0.2, 0.4,
0.6, 0.8, 1. Where level 1 means no cooperation at all and level 5 means
close cooperation, we write these evaluation results as \(t_{ij}\),
where. By scholar embedding, the vector representation of our scholars,
and then we calculated the cosine similarity \(s_{ij}\) between the
scholar vectors as follows.
\begin{equation}\label{eq:s}
    s_{ij} = \frac{\vec{A_i} \cdot \vec{A_j}}{{\mid \vec{A_i} \mid} \cdot {\mid \vec{A_j} \mid}}
\end{equation}
where \(s_{ij}\) denotes the similarity between scholar \(i\) and
scholar \(j\), \(\vec{A_i}\) denotes the vector of scholar \(i\), and
\(\vec{A_j}\) denotes the vector of scholar \(j\).

According to the above method, the cosine similarity \(\vec{s_i}\) of
scholar \(i\) and all other scholars and the evaluation result
\(\vec{t_{i}}\) are obtained in the same way, and finally we compare
\(\vec{s_i}\) with \(\vec{t_i}\).

We choose cosine similarity as the evaluation method, that is, the cosine similarity between the similarity derived by the model and the similarity annotated by manual is used as the evaluation index. The closer to 1 indicates the more accurate the model is, and we write this index as accuracy:
\begin{equation}\label{eq:Accuracy}
    Accuracy = \frac{\vec{s_i} \cdot \vec{t_{i}}}{{\mid \vec{s_i} \mid} \cdot {\mid \vec{t_{i}} \mid}}
\end{equation}

\hypertarget{results}{%
\subsection{Results}\label{results}}

\begin{figure}[htbp]
  \centering
  \includegraphics[width=0.8\linewidth]{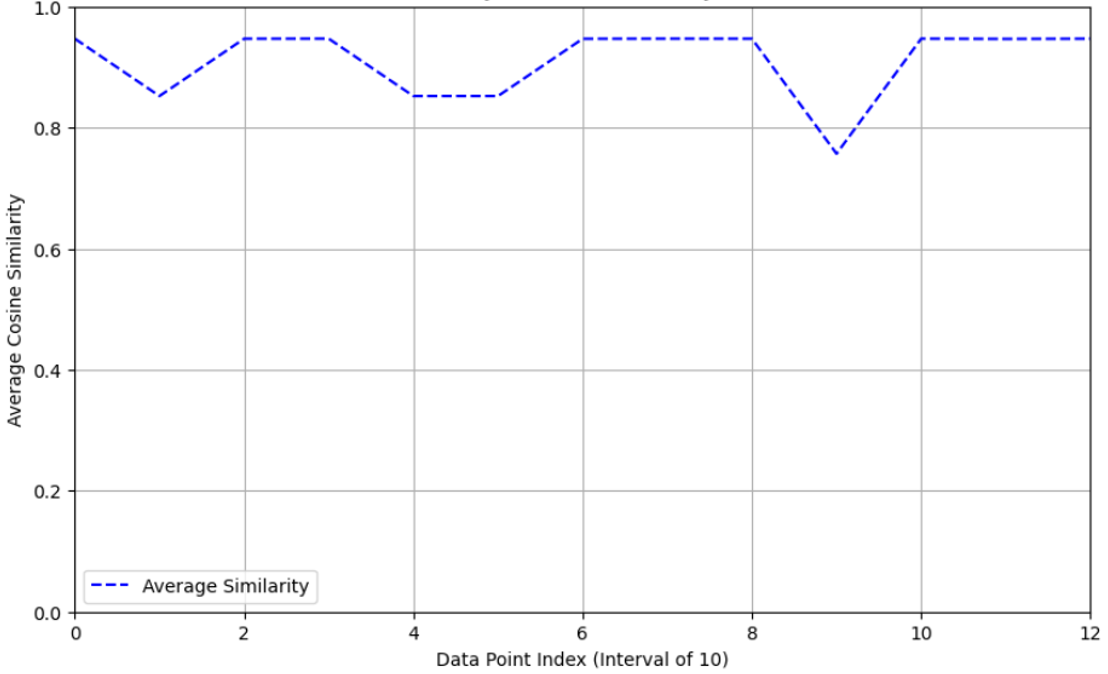}
  \caption{Accuracy Results for the Scholar Similarity Task.}
  \Description{Accuracy Results for the Scholar Similarity Task.}
  \label{fig:Accuracy} 
\end{figure}

We show the similarity between the two groups of data in Figure \ref{fig:Accuracy}. We chose 10 scholars as steps to compare \(\vec{s_i}\) with \(\vec{t_i}\), and at the same time visually observed the similarity fluctuation of the two groups of data. As you can see, the similarity of the two sets of data fluctuates within a good range, indicating that the scholar embedding we have done accurately reflects the cooperative relationship between scholar.

The final result of accuracy of scholar similarity task was calculated as
\textbf{90.91\%}, and this result well proves the feasibility of our
scheme.

\hypertarget{model analysis: selection of lambda}{%
\subsection{Model Analysis: Selection of \texorpdfstring{$\lambda$}{lambda}}\label{model analysis lambda}}

\begin{figure}[htbp] 
  \centering 
  \includegraphics[width=0.8\linewidth]{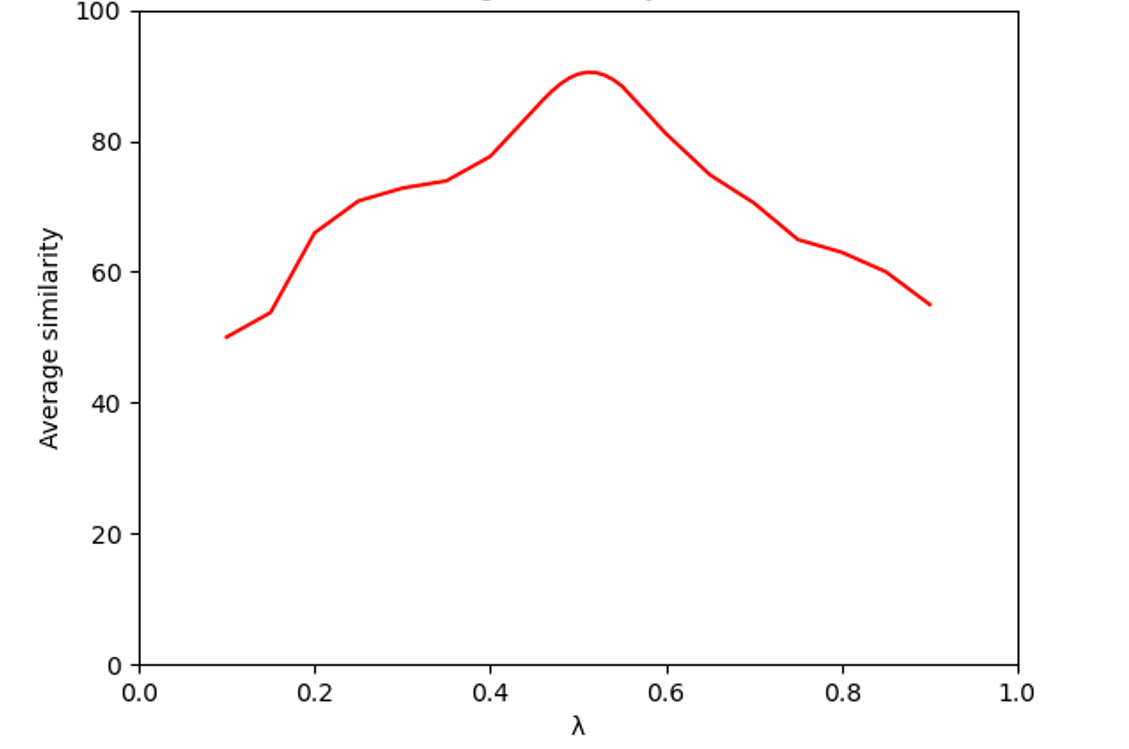}
  \caption{The Relationship between Average Similarity and $\lambda$.} 
  \Description{The Relationship between Average Similarity and $\lambda$.}
  \label{fig:Average} 
\end{figure}

In Figure \ref{fig:Average}, we show the effect of different choices of $\lambda$ in Equation \ref{eq:E} on the overall model accuracy. 
It can be seen that the accuracy increases with increasing
\(\lambda\) at the beginning. This is because when the value of
\(\lambda\) is larger, the scholar's vector is influenced by its own
contribution degree increases, which is more reflective of the scholar's
importance in that paper and makes the evaluation more accurate.
However, after the value of \(\lambda\) reaches a certain level, the
accuracy starts to decrease. This is because when the value of
\(\lambda\) is larger, the scholar's vector is influenced by the
influence of the paper increases, but this influence is determined
by the number of citations of the paper, and the number of citations of
the paper is influenced by many factors, such as the publication time of
the paper, the topic of the paper, etc. All these factors affect the
number of citations of the paper. Therefore, when the value of
\(\lambda\) reaches a certain level, the scholar's vector will be
affected by many irrelevant factors, making the evaluation inaccurate.
Therefore, we choose the value of \(\lambda\) as 0.56, a value that
enables the accuracy of the model to reach its maximum.

\hypertarget{model analysis: selection of word vector dimensions}{%
\subsection{Model Analysis: Selection of Word Vector Dimensions}\label{model analysis: selection of word vector dimensions}}

This section presents the results of the model analysis \cite{8318770}, focusing on the
selection of the most appropriate dimensionality for the word vectors in
the GloVe pre-trained model. For the analysis, we evaluate the model
using different word vector dimensions (from 50 to 300). We compare the
accuracy of scholar similarity task with different dimensions.
Table \ref{tab:addlabel} shows that the accuracy of the model for the scholar
similarity task increases and then decreases as the dimensionality of
the word vector increases, and the highest accuracy is achieved at 100
dimensions. We attribute this improvement in accuracy to the fact that a
100-dimensional scholar vector does not reflect more features of the
scholar. Therefore, we choose 100 dimensions as the dimensionality of the
word vector.
\begin{table}[htbp]
  \centering
  \caption{Accuracy of Scholar Similarity Task.}
    \begin{tabular}{cc}
    \toprule
    \multicolumn{1}{l}{dimension} & \multicolumn{1}{l}{accuracy} \\
    \midrule
    50d   & 85.16\% \\
    100d  & 90.91\% \\
    200d  & 83.45\% \\
    300d  & 80.76\% \\
    \bottomrule
    \end{tabular}%
  \Description{Accuracy of Scholar Similarity Task.}
  \label{tab:addlabel}%
\end{table}%

\newpage
\hypertarget{conclusion}{%
\section{Conclusion}\label{conclusion}}

In this paper, we propose a global semantic information and social influence
based scholar embedding model for scholar representation in the form of vector.
In our experiments, we applied the author embedding model to 126 scholars of BUPT,
obtained their respective vector representations, and calculated the 
author similarity (cosine similarity) between each pairs. We also invited the organization's scholar evaluation experts to evaluate the pairwise collaborations of 126 scholars and align them with our obtained author similarity. The experiments showed that the computed scholar vector 
achieved 90.91\% accuracy in the scholar similarity task. In other words, 
the author vector well characterizes the scholarly achievements and 
collaborations of interdisciplinary scholars, and our author embedding 
model successfully quantifies the scholars.

Our author embedding model opens a new door for future interdisciplinary
collaborative network discovery and recommendation research. It solves
one of the most fundamental scholar quantification and representation task,
and lays the foundation for our subsequent work, including
interdisciplinary scholar evaluation, interdisciplinary collaborative
network discovery-evaluation-recommendation, and more innovative applications.

\bibliographystyle{ACM-Reference-Format}
\bibliography{reference}

\end{document}